\documentclass[conference]{IEEEtran}
\IEEEoverridecommandlockouts
\usepackage{cite}
\usepackage{amsmath,amssymb,amsfonts}
\usepackage{algorithmic}
\usepackage{graphicx}
\usepackage{textcomp}
\usepackage{url}
\usepackage{xcolor}
\usepackage{svg}

\def\BibTeX{{\rm B\kern-.05em{\sc i\kern-.025em b}\kern-.08em
    T\kern-.1667em\lower.7ex\hbox{E}\kern-.125emX}}
\begin{document}

\title{Realism in Action: Anomaly-Aware Diagnosis of Brain Tumors from Medical Images Using YOLOv8 and DeiT
}

\author{
\IEEEauthorblockN{1\textsuperscript{st} Seyed Mohammad Hossein Hashemi}
\IEEEauthorblockA{\textit{Department of CS \& IT} \\
\textit{Institute for Advanced Studies
in Basic Sciences (IASBS)}\\
Zanjan, Iran \\
mh.hashemi@iasbs.ac.ir}
\and
\IEEEauthorblockN{2\textsuperscript{nd} Leila Safari}
\IEEEauthorblockA{\textit{Department of Computer Engineering} \\
\textit{University of Zanjan (ZNU)}\\
Zanjan, Iran \\
lsafari@znu.ac.ir}
\and
\IEEEauthorblockN{3\textsuperscript{rd} Mohsen Hooshmand}
\IEEEauthorblockA{\textit{Department of CS \& IT} \\
\textit{Institute for Advanced Studies
in Basic Sciences (IASBS)}\\
Zanjan, Iran \\
mohsen.hooshmand@iasbs.ac.ir}
\and
\IEEEauthorblockN{4\textsuperscript{th} Amirhossein Dadashzadeh Taromi}
\IEEEauthorblockA{\textit{Department of CS \& IT} \\
\textit{Institute for Advanced Studies
in Basic Sciences (IASBS)}\\
Zanjan, Iran \\
dadashzadeh@iasbs.ac.ir}
}

\maketitle

\begin{abstract}
Reliable diagnosis of brain tumors remains challenging due to low clinical incidence rates of such cases. However, this low rate is neglected in most of proposed methods.  We propose a clinically inspired framework for anomaly-resilient tumor detection and classification. Detection leverages YOLOv8n fine-tuned on a realistically imbalanced dataset (1:9 tumor-to-normal ratio; 30,000 MRI slices from 81 patients). In addition, we propose a novel Patient-to-Patient (PTP) metric that evaluates diagnostic reliability at the patient level. Classification employs knowledge distillation: a Data Efficient Image Transformer (DeiT) student model is distilled from a ResNet152 teacher. The distilled ViT achieves an F1-score of 0.92 within 20 epochs, matching near-teacher performance (F1=0.97) with significantly reduced computational resources. This end-to-end framework demonstrates high robustness in clinically representative anomaly-distributed data, offering a viable tool that adheres to realistic situations in clinics.
\end{abstract}


\section{Introduction}
\label{sec:introduction}

Brain tumors pose severe health risks where delayed diagnosis critically impacts survival outcomes \cite{tandel2019review}. Magnetic Resonance Imaging (MRI) serves as the clinical gold standard for detection due to its superior soft-tissue resolution \cite{augustine2021imaging}, yet manual interpretation struggles with the extreme rarity of tumors (\textless0.1\% incidence \cite{JHhowcommon}) amidst vast normal data \cite{popuri20123d}. This \textit{clinical imbalance}, compounded by high tumor heterogeneity \cite{kang2021mri}, undermines traditional computer-aided diagnosis systems.

Existing deep learning approaches—including GAN-augmented CNNs \cite{ahmad2022brain}, attention mechanisms \cite{abdusalomov2023brain}, and Vision Transformers (ViTs) \cite{asiri2023exploring}—often rely on semi-balanced datasets misrepresenting real-world scarcity. While achieving high image-level accuracy (e.g., 98.7\% for ensemble ViTs \cite{tummala2022classification}), they neglect patient-level diagnostic reliability and practical deployability in resource-constrained settings \cite{sharif2021decision,asiri2023exploring}.

To bridge this gap, we introduce an \textit{anomaly-aware} framework comprising:
\begin{enumerate}
	\item \textbf{Clinically representative data}: Curated NBML dataset (81 patients; 30,000 MRI slices) preprocessed to enforce 1:9 tumor-to-normal ratio both at slice and patient level.
	\item \textbf{Patient-level evaluation}: Novel \textbf{PTP metric} assessing diagnostic reliability across full patient studies.
	\item \textbf{Efficient two-stage architecture}:
	\begin{itemize}
		\item \textit{Detection}: YOLOv8n fine-tuned for robust localization in imbalanced data.
		\item \textit{Classification}: Knowledge distillation via DeiT \cite{touvron2021training}, transferring ResNet152 insights to compact ViT.
	\end{itemize}
\end{enumerate}
Our approach optimizes both accuracy (PTP-F1=1.0; classification F1=0.92) and computational efficiency for clinical deployment.

Section II is a literature review on related work in this field. Section III describes the dataset used in this study. Section V presents the results we achieved, and Section VI concludes the article.

\section{Related Work}
\label{sec:Related Work}

Recent hardware advances, particularly GPUs, have demonstrated deep learning as the dominant paradigm for brain tumor diagnosis \cite{shaik2022multi,alanazi2022brain}. Representative works from key methodological approaches demonstrate both progress and persistent challenges:

Ahmad \textit{et al.} \cite{ahmad2022brain} pioneered generative data augmentation using VAE-GAN hybrids, boosting ResNet50 accuracy to 96.25\%. However, such methods incur prohibitive computational costs that limit clinical utility. Sharif \textit{et al.} \cite{sharif2021decision} exemplified feature selection techniques with their EKbHFV-MGA framework, achieving 95\% accuracy but introducing complexity that impedes cross-dataset generalization. Vision Transformer innovations are well-represented by Asiri \textit{et al.} \cite{asiri2023exploring} (FT-ViT) and Tummala \textit{et al.} \cite{tummala2022classification} (ViT ensembles), reaching up to 98.7\% accuracy at the cost of extreme computational demands. For detection tasks, Abdusalomov \cite{Abdusalomov2023Aug} demonstrated attention-enhanced YOLOv7 with CBAM/SPPF+ layers achieving 99.5\% accuracy, though crucially evaluated on balanced datasets that misrepresent clinical reality.

While these representative works achieve high image-level accuracy, they collectively neglect three critical clinical requirements: 1) patient-level diagnostic reliability; 2) performance under true incidence rates (\textless0.1\% tumors); and 3) computational constraints of medical environments. Our framework addresses these gaps through novel patient-centric evaluation (PTP metric) and resource-optimized architecture design.

\section{Data and Resources}
\label{sec:data}
\subsection{NBML Dataset}

The dataset from the National Brain Mapping Lab (NBML) which we used for detection includes MRI slices captured through T1-weighted, T2-weighted, and diffusion-weighted sequences, as well as other imaging modalities such as PET and CT scans. It is important to note that this dataset remains privately held, with all associated rights and credits attributed to the Iranian Brain Mapping Biobank (IBMB).

\subsection{Kaggle Dataset}
For tumor classification, we combined and preprocessed two public sources namely Kaggle Brain Tumor MRI Dataset (accessed on August 2023) and Figshare dataset\cite{cheng2017brain}.

Our custom augmentation module generated variations to increase sample diversity and it was partitioned into training, validation, and benchmark sets for rigorous evaluation.

\subsection{Computational Resources}
Experiments used an NVIDIA GTX 1650 GPU locally and Google Colab's T4 GPU. We employed PyTorch 2.0.1+cu117/Python 3.9.7, with ViT from Wang's GitHub \cite{vitpytorch} and TorchVision's ResNet152.

\section{Proposed method}

Accurate and realistic brain tumor diagnosis involves two distinct goals: 1) detecting tumors within a predominantly normal dataset, and 2) identifying unusual brain tissues and their types from unique characteristics in noisy scenes (e.g., shape and suspicious tissue placement).

\label{sec:data}
\begin{figure*}[htbp]
	\centering    \includegraphics[width=2\columnwidth]{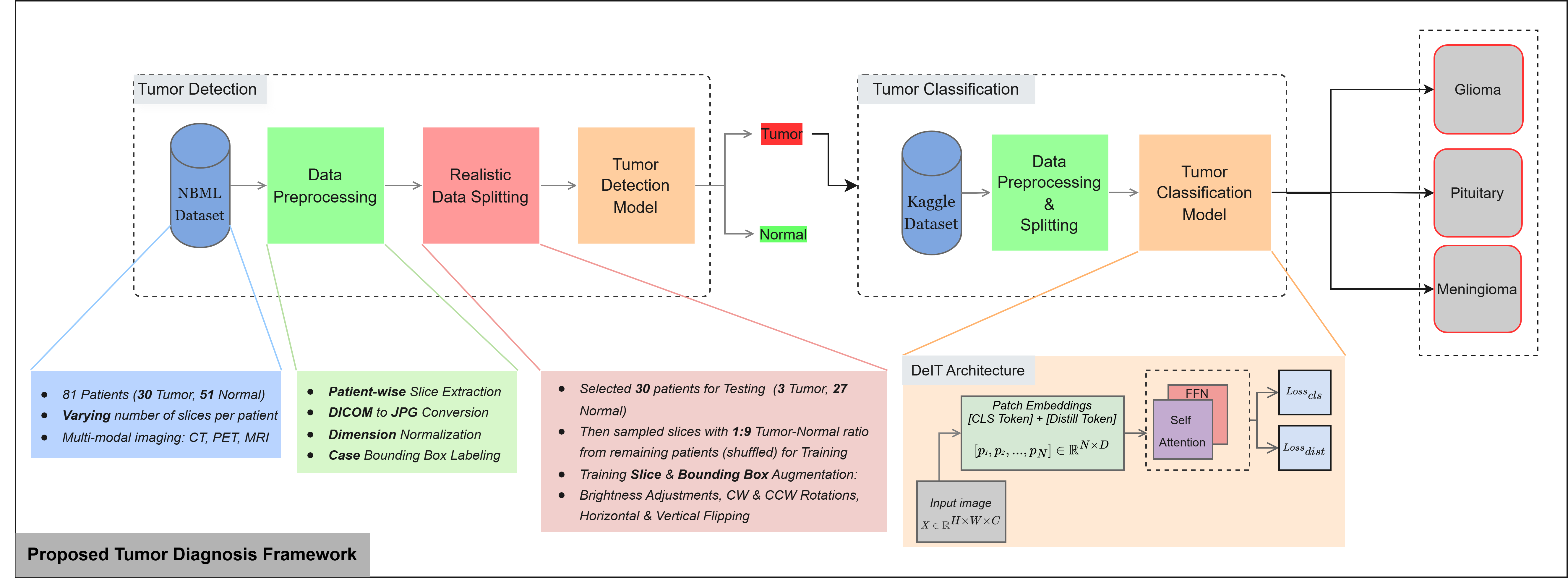}
	\caption{Proposed Framework}
	\label{fig:framework}
\end{figure*}

The tumor detection, focuses on training a model that is resilient to \textit{anomaly-distributed} populations and can accurately detect brain tumors in various imaging modalities. 

The tumor classification, involves designing a custom DeiT model and training it to classify brain tumors in three classes of Meningioma, Pituitary, and Glioma.

\subsection{Tumor detection}
This phase utilizes the NBML dataset with 81 patients (30 tumor, 51 normal) to simulate clinical imbalance through specialized preprocessing.

\subsubsection{\textbf{Rationale for Choosing YOLOv8}}
We selected YOLOv8n for tumor detection based on three key advantages:
\begin{itemize}
	\item \textit{Performance in complex scenarios}: Architectural enhancements (advanced backbone/neck modules) improve detection in noisy medical images \cite{Solawetz2023Dec}.
	\item \textit{Generalization capability}: Effective transfer learning compensates for limited annotated data \cite{Ragab2024Apr}.
	\item \textit{Computational efficiency}: Optimized for deployment in resource-constrained clinical environments.
\end{itemize}
The model was fine-tuned for our detection task and evaluated using both standard metrics and our novel PTP framework.

\subsubsection{\textbf{Data Preprocessing}}
The essence of our data preparation pipeline centers on the careful pre-processing of our data set to closely mirror real-world scenarios. 
In the context of brain tumors, the United States typically reports \cite{SEERcancer} incidence rate less than 0.1. Given the limitations stemming from our limited dataset for the detection phase and the absence of comprehensive external data sources, we exercised caution by adopting a conservative estimate of a 0.1 incidence rate.

\subsubsection{\textbf{Realistic splitting}}

To maintain clinically representative data distribution, we partitioned the dataset into training and testing sets while preserving our target 1:9 tumor-to-normal ratio. For the training set, we ensured nine randomly selected normal images were paired with each tumor image. This approach trains the model to recognize robust features across diverse patient scenarios rather than individual cases.

For the testing set, we selected 30 complete patient studies: 27 normal cases and 3 tumor cases (10\% incidence), preserving all associated images per patient. This presented significant challenges due to:
\begin{itemize}
	\item Variable image counts across patients
	\item Complex directory structures with modality-specific subfolders (PET/CT/MRI types)
	\item Initial dataset imbalance versus target distribution
	\item Tumor-free slices in tumor patient folders
\end{itemize}

Here is our systematic approach:

\begin{enumerate}
	\item DICOM to JPG conversion (540$\times$540) using MicroDicom.
	
	\item Patient directory compression to ZIP archives to:
	\begin{itemize}
		\item Efficiently quantify total image volume.
		\item Avoid recursive directory traversal of nested modality folders.
	\end{itemize}
	
	\item Testing set selection by ZIP size:
	\begin{itemize}
		\item 27 normal patients: Smallest ZIPs.
		\item 3 tumor patients: Smallest uncleaned ZIPs.
	\end{itemize}
	
	\item Training set construction:
	\begin{itemize}
		\item $\sim$1.4k tumor-indicative images (cleaned cases).
		\item 12,000 normal images (1:9 tumor-to-normal ratio).
	\end{itemize}
	
	\item Augmentation implementation:
	\begin{itemize}
		\item Brightness adjustments,
		\item Vertical/horizontal flips,
		\item Bounding box-preserving rotations.
	\end{itemize}
\end{enumerate}

\subsubsection{\textbf{Proposed Evaluation Metrics}}

Standard image-level metrics were employed alongside our novel Patient-to-Patient (PTP) framework to evaluate model effectiveness. Precision, Recall, and F1-score are defined as:

\begin{equation}Precision=TP/(TP+FP)\label{eq},\end{equation}
\begin{equation}Recall=TP/(TP+FN)\label{eq},\end{equation}
\begin{equation}F1=2 \times(Precision\times Recall)/(Precision+Recall)\label{eq}\end{equation}

where $TP$ denotes True Positives, $TN$ True Negatives, $FP$ False Positives, and $FN$ False Negatives.

To address clinical needs in imbalanced settings, we developed the PTP metric framework. This approach processes all images per patient directory, computes a Patient-Specific Tumor Threshold (PSTT) as:

\begin{equation}
	\text{PSTT} = \frac{\text{\# Tumor-indicative images}}{\text{\# Total images}}
\end{equation}

and classifies the case as tumor-positive if $\text{PSTT} > \text{GTT}$. Based on these definitions, four evaluation metrics are derived:

\textbf{PTP-Accuracy} measures overall patient classification correctness, calculated as the proportion of correctly classified patients. 

\textbf{PTP-Recall} quantifies sensitivity for tumor patients, representing the fraction of actual tumor patients correctly identified. 

\textbf{PTP-Precision} assesses positive predictive value by measuring the proportion of correctly identified tumor patients among all patients classified as tumor-positive. 

\textbf{PTP-F1} provides a balanced measure through the harmonic mean of PTP-Precision and PTP-Recall:
\begin{equation}
	\text{PTP-F1} = \frac{2 \times \text{PTP-Precision} \times \text{PTP-Recall}}{\text{PTP-Precision} + \text{PTP-Recall}}
\end{equation}

This framework is exclusively deployed for tumor detection evaluation, while classification phase assessment utilizes standard accuracy metrics.

\subsubsection{\textbf{General Tumor Threshold}}

\begin{figure}[htbp]
	\centering
	\includegraphics[width=\columnwidth]{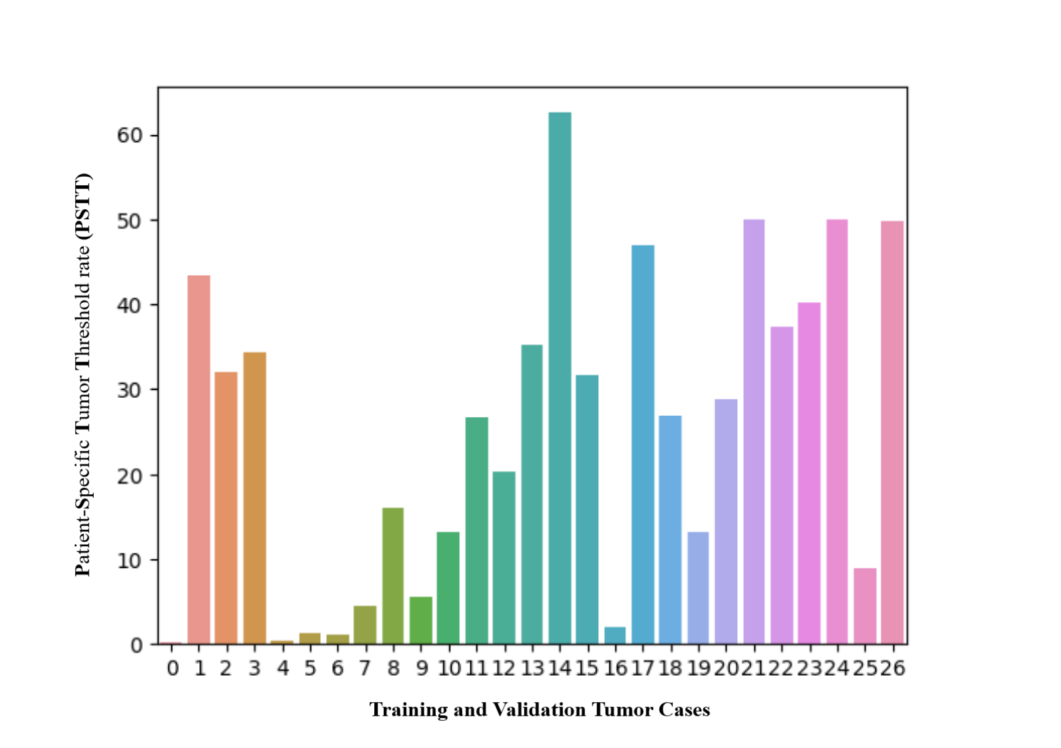}
	\caption{PSTT, computed for each patient, represents the proportion of images depicting tumors within the entire image set for that patient.}
	\label{fig1}
\end{figure}

The General Tumor Threshold (GTT) is the minimum percentage of tumor-indicative images required within a patient's complete scan set to classify them as tumor-positive. We determined this critical threshold through careful analysis of our training and validation data.

We processed all the patients scans within the training and validation sets through our detection model and calculated the value of GTT for each patient. After an exploratory analysis of the PSTT (Fig 2) values among the both Normal and Tumor cases, we estimated the value of the GTT to be at least be 0.04\%. This value is calculated from the average of the first quartile and the median values of tumor-indicative distribution.

\subsection{Tumor classification}

In this step, We employed Knowledge Distillation (KD) \cite{hinton2015distilling} to train a lightweight Vision Transformer (ViT) using our classification dataset, with ResNet152 as the teacher model. KD enables the student model (ViT) to learn from the teacher's complete output distribution (soft labels), not just final predictions. This compresses knowledge from large models into efficient architectures while boosting performance.

\subsubsection{\textbf{Justification for Choosing DeiT}}
 
In this study, the DeiT \cite{touvron2021training} model was selected as the backbone for the classification phase due to several important factors that make it particularly suitable for our use case:

\begin{enumerate}
	\item \textit{Data Efficiency}: DeiT excels well with limited medical imaging data
	\item \textit{Performance Balance}: It maintains high accuracy while reducing model size
	\item \textit{Computational Advantage}: Distillation from ResNet152 (Strong Teacher) enables:
	\begin{itemize}
		\item Faster training than standard ViTs
		\item Lower inference costs
		\item Near-teacher performance (least compromise)
	\end{itemize}
\end{enumerate}

\subsection{Vision Transformer}
Our pipeline employs the Vision Transformer (ViT) \cite{dosovitskiy2020image}, adapting language processing principles to images. Key processing stages:
\begin{itemize}
	\item \textit{Patch creation}: Splits input ($\mathbb{R}^{H\times W\times C}$) into $N$ patches of size $P\times P\times C$ where $N = HW/P^2$
	\item \textit{Embedding}: Linear projection with added positional encodings
	\item \textit{Classification token}: Learnable [CLS] embedding for final prediction
	\item \textit{Encoder}: Transformer blocks alternating multi-head self-attention and MLP operations
\end{itemize}
This architecture provides global context modeling \cite{vaswani2017attention}, crucial for tumor classification \cite{asiri2023exploring}.

\section{Results}
\label{sec:Results}

This section elaborates on the experiments and the results we achieved from deploying the proposed pipeline.

\subsection{Tumor Detection Results}

The initial step in this phase was data pre-processing. After conducting a comprehensive data prepration, we loaded the YOLOv8n pre-trained weights, tailored its hyper-parameters and fine-tuned it on our detection dataset.

Due to computational constraints, we selected YOLOv8n (Nano) and used larger batch sizes to accelerate training. While advanced versions (YOLOv8m/L) may improve results, they require significantly more resources and longer training times.

\begin{table}[htbp]
	\caption{{YOLOv8n Model Configuration}}
	\centering
	\begin{tabular}{|c|c|c|c|c|c|}
		\hline
		\textbf{Opt} &
		\textbf{Sched} &
		\textbf{lr0} &
		\textbf{lrf} &
		\textbf{AMP} &
		\textbf{Epochs} \\
		\hline
		SGD & CosLR & 0.01 & 0.00001 & False & 40\\
		\hline
	\end{tabular}
	\par
	\bigskip
	Opt: Optimizer, Sched: Scheduler, lr0: Initial learning rate, lrf: Final learning rate, AMP: Automatic Mixed Precision.
	\label{label}
\end{table}

We trained the mentioned model for 40 epochs, and the evaluation results indicated (Table II) that the model is highly accurate in detecting tumor-confiscated images, and despite being agile and super lightweight with only 3.2M parameters, it does have a reliable performance.

\begin{table}[htbp]
	\caption{{YOLOv8n Model Evaluation Results}}
	\centering
	\begin{tabular}{|c|c|c|c|c|}
		\hline
		\textbf{Class} & \textbf{Precision} & \textbf{Recall} & \textbf{F1} & \textbf{Support}  \\
		\hline
		Tumor & 0.99 & 0.96 & 0.97 & 1905 \\
		\hline
		Normal & 0.99 & 0.99 & 0.99 & 20750 \\
		\hline
		\textbf{AVG} & 0.99 & 0.975 & 0.98 & 22655 \\
		\hline
	\end{tabular}
	\label{label}
\end{table}


\begin{figure*}[htbp]
	\centering
	\includegraphics[width=\textwidth]{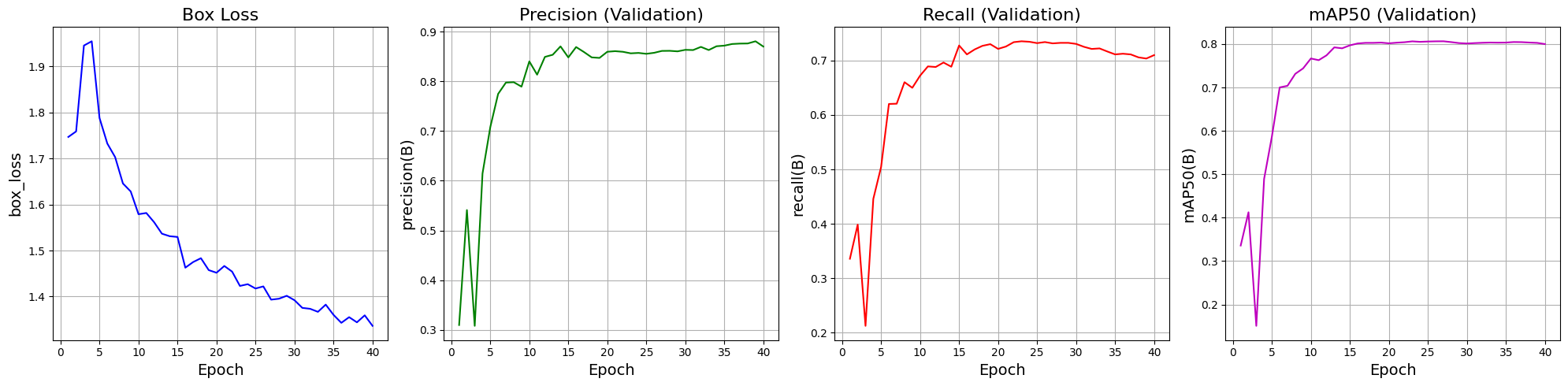}  
	\caption{Validation Performance Metrics and Training Loss Over Epochs for the YOLOv8n Detection Model.}
	\label{fig2}
\end{figure*}

As for the patient-level assessment, the detection model achieved perfect scores for both tumor and normal cases, with an PTP F1-score of 1.0 across the testing population. This highlights the model’s ability to generalize well to real-world clinical scenarios, where detecting tumors in anomaly-distributed data is crucial.

Furthermore, the performance of the detection model, as demonstrated in Fig. 3, provides a clear indication of its robustness. Throughout training, the validation metrics show consistent improvement, while the training loss decreases steadily. This parallel progression between training and validation suggests a well-balanced model that is neither underfitting nor overfitting. Overfitting manifests as a divergence between training and validation results—where training loss decreases but validation performance plateaus or worsens.

\subsection{Tumor Classification Results}

\begin{table*}[htbp]
	\caption{{DeiT hyper-parameters tuning experiments}}
	\centering
	\begin{tabular}{|c|c|c|c|c|c|c|c|c|}
		\hline
		\textbf{No.} & 
		\textbf{Hard Distillation} & 
		\textbf{Temperature} &
		\textbf{Depth} & 
		\textbf{Patch Size} &
		\textbf{Dimension} &
		\textbf{Attention Head} &
		\textbf{MLP Dim} &
		\textbf{Val-Accuracy} \\
		\hline
		1 & False (Default) & 2 (Default) & 4 (Default) & 24 (Default) & 256 (Default) & 16 (Default) & 128 (Default) & 81.91 \\
		\hline
		2 & True & 2 (Default) & 4 (Default) & 24 (Default) & 256 (Default) & 16 (Default) & 128 (Default) & 84.74 \\
		\hline
		3 & True & 1 & 4 (Default) & 24 (Default) & 256 (Default) & 16 (Default) & 128 (Default) & 83.22 \\
		\hline
		4 & True & 9 & 4 (Default) & 24 (Default) & 256 (Default) & 16 (Default) & 128 (Default) & 81.69 \\
		\hline
		5 & True & 3 & 6 & 24 (Default) & 256 (Default) & 16 (Default) & 128 (Default) & 82.35 \\
		\hline
		6 & True & 3 & 2 & 32 & 256 (Default) & 16 (Default) & 128 (Default) & 85.40 \\
		\hline
		7 & True & 3 & 2 & 24 & 256 (Default) & 16 (Default) & 128 (Default) & 86.05 \\
		\hline
		8 & True & 3 & 2 & 24 & 1024 & 16 (Default) & 128 (Default) & 68.19 \\
		\hline
		9 & True & 3 & 2 & 24 & 128 & 16 (Default) & 128 (Default) & 85.40 \\
		\hline
		10 & True & 3 & 2 & 24 & 512 & 16 (Default) & 128 (Default) & 74.29 \\
		\hline
		11 & True & 3 & 2 & 24 & 128 & 64 & 128 (Default) & 88.67 \\
		\hline
		12 & True & 3 & 2 & 24 & 128 & 64 & 256 & 88.45 \\
		\hline
		13 & True & 3 & 2 & 24 & 128 & 64 & 2048 & 87.58 \\
		\hline
		14 & True & 3 & 2 & 24 & 128 & 64 & 512 & 89.76 \\
		\hline
		
	\end{tabular}
	\label{label}
\end{table*}

Our classification phase employed Knowledge Distillation (KD) with DeiT architecture. We constructed an augmented dataset from the Figshare source using our custom augmentation module, then fine-tuned the ResNet152 teacher model to leverage its robust feature extraction capabilities. During distillation, ResNet152 transferred rich feature representations to the DeiT student model, enabling efficient training from scratch.

We proceeded to tune the hyperparameters of the DeiT model by iteratively exploring 14 different architectural variations to find the optimal settings. The specifics of these experiments are detailed in Table III.

Then we evaluated the optimal DeiT configuration on 461 test images across three tumor classes and achieved:
\begin{itemize}
	\item \textbf{Pituitary}: F1-Score = 0.97
	\item \textbf{Glioma}: F1-Score = 0.93
	\item \textbf{Meningioma}: F1-Score = 0.82
\end{itemize}
 
The results in Table IV indicates that the student model despite having access to \textit{relatively small} size dataset and being trained for \textit{limited number} of epochs, is still effective in learning the distinguishing features of each tumor type. Also the Meningioma class seems to have a data integrity problem and the original data sources seems to lack the data diversity required for training a reliable model on them. This issue could be solved by SMOTE or weighted loss functions and perhaps data augmentation, however the main objective here was to prove that even in these scenarios the student model still performs pretty good.

As indicated in Table IV, student model achieved a competitive F1-Score of 0.92 compared to the ResNet152 teacher's 0.97 (Table V). This performance difference primarily stems from the teacher's training on heavily augmented data, which enhanced feature generalization but required substantial computational resources. In contrast, the DeiT student was trained for only 20 epochs without augmentation using minimal resources. 

\begin{table}[htbp]
	\caption{{Distilled Student Classifier Test Results}}
	\centering
	\begin{tabular}{|c|c|c|c|c|}
		\hline
		\textbf{Tumor Class} & \textbf{Precision} & \textbf{Recall} & \textbf{F1} & \textbf{Support}  \\
		\hline
		Meningioma & 0.82 & 0.82 & 0.82 & 107 \\
		\hline
		Glioma & 0.95 & 0.92 & 0.93 & 214 \\
		\hline
		Pituitary & 0.95 & 0.99 & 0.97 & 140 \\
		\hline
		\textbf{Weighted AVG} & \textbf{0.92} & \textbf{0.92} & \textbf{0.92} & \textbf{461}  \\
		\hline
	\end{tabular}
	\label{label}
\end{table}

\begin{table}[htbp]
	\caption{{Teacher Classifier Test Results}}
	\centering
	\begin{tabular}{|c|c|c|c|c|}
		\hline
		\textbf{Tumor Class} & \textbf{Precision} & \textbf{Recall} & \textbf{F1} & \textbf{Support}  \\
		\hline
		Meningioma & 0.92 & 0.91 & 0.91 & 107 \\
		\hline
		Glioma & 0.99 & 0.97 & 0.98 & 214 \\
		\hline
		Pituitary & 0.94 & 0.97 & 0.96 & 140 \\
		\hline
		\textbf{Weighted AVG} & \textbf{0.97} & \textbf{0.97} & \textbf{0.97} & \textbf{461}  \\
		\hline
	\end{tabular}
	\label{label}
\end{table}

Crucially, both models were evaluated on an identical holdout test set (15\% benchmark allocation), preventing data leakage while ensuring fair comparison. This confirms DeiT's efficiency for clinical deployment.

\section{Conclusion}

This work presented a novel framework for the detection and classification of brain tumors. To realistically simulate the low incidence rate brain tumor clinical diagnosis scenarios, an extensive, meticulous data preprocessing steps were applied. 

As for the detection, we introduced a new set of performance metrics namely PTP-metrics with a focus on capturing the performance in clinical scenarios. Further, we trained YOLOv8 for a few epochs and achieved near to perfect results, indicating a anomaly robust performance.

Furthermore, in the classification we distilled an student model from ResNet152 teacher using DeiT architecture and achieved comparative performance.

This work is the first in literature that introduced a close to clinical framework to capture the tumor detection performance at patient level scope rather than inidvidual slices, hence further refinement of the GTT and the extension of PTP-like metrics to classification task remains an area for potential future investigation.

\label{sec:conc}

\subsection*{Code Availability}
The implementation code is available at:\\ \url{https://github.com/MHosseinHashemi/NBML_BrTD}

\bibliographystyle{IEEEtran}
\bibliography{main}
\end{document}